\documentclass[9pt]{article}
\usepackage{graphicx}
\usepackage{epstopdf}
\usepackage{dcolumn}
\usepackage{bm}

\title{Speakable and unspeakable  in cosmology:  dark matter vs. gravitational self energies. Hubble's constant, the cosmological term and all that.}

\begin{document}

\author {Paolo Christillin \\
Dipartimento di Fisica, \\
Universit\`a di Pisa\\
and \\  I.N.F.N. , Sezione di Pisa}

\maketitle

\begin {abstract}  

The inadequacy of the present cosmological picture is underlined.  The central issue of  energy and particles-photons number conservation is addressed.   It is shown that  consideration of gravitational self energy is paramount  both for matter and for radiation to bring present data estimates of matter and radiation density and the radius of the universe towards agreement with the Planck scale quantities from which it should have consistently evolved.   Particle creation  is proven to play a fundamental role in the evolution of the Universe. It is argued that we might be living inside an expanding black hole.

\end {abstract}

\


\

\section{Introduction}\label{}

Within the present theoretical framework the Planck units 

\begin{equation}\label{tP}
t_{P} \simeq (\hbar G/c^5)^{1/2}\simeq 10^{-44}  sec
\end{equation}

\

\begin{equation}\label{tP}
R_{P} \simeq (\hbar G/c^3)^{1/2}\simeq 10^{-33}  cm
\end{equation}

\

\begin{equation}\label{tP}
M_{P} \simeq (\hbar c/G)^{1/2} \simeq 10^{19}  m_{p}
\end{equation}

\

( where  $m_{p}$ stands for the proton mass), with the obvious relations 

\

\

\begin{equation}\label{RP}
E_{P} R_{P} = \hbar c    
\
,
\\\
E_{P} t_{P}  = \hbar 
\
,
\
R_{P}/ t_{P}  = c
\end{equation}

are believed to represent the primordial scales, from which, with a big bang, our Universe should have evolved. They correspond to a regime where all interactions should have unified, combining gravity (G) with special relativity (c) and quantum mechanics ($\hbar$),  $G \simeq c \simeq \hbar \simeq  1$.

\
The two experimental pieces of evidence in support for such a picture are the $2.7 K^{o}$  cosmic background radiation (CBR)  (\cite{Penzias}) eand the red shift of distant galaxies, whose recession velocity at a distance R from us is given by the famous Hubble's law  (\cite{Hubble})

\begin{equation}\label{tP}
v = H_{o} R 
\end{equation}

where $H_{o} \simeq 2.5 \times 10^{-18} \; sec^{-1}$.

This more or less determines the radius of our present (visible) universe $ R_{U} = c/H_{o}  \simeq 10^{26} m$. 
Hubble's constant is fundamental in determining the critical density $\rho_{c} $

\begin{equation}\label{tP}
\rho_{c}  \simeq 3 H_{o}^2/ (8 \pi G) \sim  
10^{-26} kg/m^3 \simeq 5 m_{p} /m^3 
\end{equation}

which is thought to determine the fate of our universe.

Indeed, assuming homogeneity, according to whether the present density (matter + radiation)  $\rho_{G}  $  is bigger or smaller than $\rho_{c} $ , our Universe should
eventually contract or keep expanding.

The fact that  the order of magnitude of 

\begin{equation}\label{tP}
\rho_{G}  \simeq M_{U}/(4/3  \: \pi  R_{U}^3 )\simeq .5 \times 10^{-26} kg/m^3    
\end{equation}

appears to be smaller than the critical one, unavoidably  prompts  speculations about dark matter \cite {vera} , \cite {Sidney} and dark energy. 

Let us now check the consistency of these data with former history and in particular with the supposed origin of the universe \cite{Fang}.  

\section{The  energy problem. }\label{en}

\

Let us  start by considering radiation , since its correct quantum  behaviour stays at the foundation of all of our present  theoretical understanding. 
Its present energy density, coming from the CBR, is 

\begin{equation}\label{tP}
\epsilon_{r }  \simeq (kT_{r})^4 \simeq 10^5 eV/m^3 
\end{equation}

which yields (numbers having been rounded off, estimates should be understood to be valid to within an order of magnitude) a total radiation energy for the Universe at present 

\begin{equation}\label{}
E_{r,U} \simeq  (kT_{r})^4 (R_{U})^3 \simeq 10^{74}  GeV 
\end{equation}

In addition, given the actual photon density (number  of photons per cubic meter) $n_{\gamma} \simeq 10^{9} $ ,  just statistical  fluctuations of the temperature

\begin{equation}\label{tP}
\Delta T_{r} / T_{r} = 1/3 ( \Delta n_{\gamma}/ n_{\gamma} ) = 1/ (3  \sqrt n_{\gamma} ) 
\end{equation}

entail 

\begin{equation}\label{tP}
\Delta T_{r} / T_{r}  \simeq 10^{-5}   
\end{equation}

 accounting  for the anisotropy of CMB   (traditionally  thought to provide a "standard ruler" for the geometry of space).
 
 This  gives us some confidence in the interpretation of present radiation data.

The thermodynamics of an adiabatic expansion predicts 

\begin{equation}\label{tP}
T_{r }  \simeq 1/R
\end{equation}

implying  that the total number of photons $N_{\gamma } $

\begin{equation}\label{tP}
N_{\gamma } \simeq n_{\gamma } \;  R^3   \simeq  (k T_{r }R)^3
\end{equation}

stays constant, and that therefore the {\it total}  radiation energy simply goes as $T_{r }$.  

It follows that in going backward in time

\begin{equation}\label{e'ru}
E' _{r,U}  = E_{r,U}  ( T'_{r}/T _{r} )
\end{equation}

the radiation energy in the shrinking Universe would keep on growing (primed quantities referring  to previous times) .  

Let us then move on to matter to which the black body  treatment does not apply . 

 The total number of baryons in the Universe is estimated to be $\simeq 10^{80} $.  The total energy of matter in the Universe is hence 

\begin{equation}\label{e'mu}
E_{m,U}  \simeq 10^{80} GeV  
\end{equation}

If, along standard lines, we assume baryon number conservation  

\begin{equation}\label{tP}
N_{m } \simeq n_{m } \;  R^3  
\end{equation}

(where $ n_{m}$ stands for the baryonic density which must therefore go as $1/R^3 $ ) 

and that the matter energy density be essentially given by the baryonic mass $m_{p} $ (this position, which is the {\it standard one} \cite {Fang}, \cite {Perl}, \cite {wood} , should, to a closer scrutiny, be corrected for temperature and velocity dependence of the baryons. It is however easy to see that the former is totally irrelevant up to a temperature of $10^{10} K = 1 MeV$ and that the latter even for $v=c$ would just give a factor of $\sqrt {2} $),
then  the total baryonic energy 

\begin{equation}\label{tP}
E'_{m,U}  \simeq \epsilon _{m} R^3 =   n_{m} \; m_{p}  c^2 R^3
\end{equation}

stays constant.  

The previous considerations are  used  to determine the temperature  of the passage from  the present matter dominated to radiation dominated era .

Fom Eq.s (\ref{e'ru}),( \ref{e'mu}) , by requiring $ E'_{r,U}  = E'_{m,U}$ we have 

\begin{equation}\label{tP}
  (T' _{r}/ T_{r}) = E_{m,U}/E_{r,U}
\end{equation}

i.e. $ T'_{r}  \simeq 10^6 \times 2.7 \; K  $ ,a striking $O (10^3) $  difference with respect to  the standard result $ T'_{r}  \simeq 3000 \; K$.

This already  queer scenario concerning   energy becomes even more puzzling if we try to push our picture that radiation must have been prevailing also in the earlier history,  by assuming a constant photon number .

The temperature at the Planck scale should hence be given    by

\begin{equation}\label{tP}
T_{r,P } =   T_{r,U } R_{U }/R_{P }
\end{equation}

i.e.

\begin{equation}\label{tP}
T_{r, P } \simeq 10^{61}  K  \;  \;   \;  and   \;  \; \;  E_{r,P} \simeq 10^{135}  GeV
\end{equation}

to be compared to the  Planck energy 
  $ E_{P} \simeq 10^{19} GeV  $

Translated in time the previous result would in turn  imply either a big bang time much smaller than Planck's or an age of the Universe much larger than the present estimate, both things being manifestly untenable.

How can we solve such a totally disastrous energy non conservation problem ?

{\it The point is that  we have forgotten self gravitational effects and this, in turn, allows for photons and matter creation. }  In other words the present estimate for the number of  photons and nucleons pertains only to the recent history of the Universe and the previous arguments about "particles" cannot be pushed too backward in time.

\

\section{Self energies  and particles creation. }\label{self}

Starting with radiation we must underline that its  treatment {\it at ordinary conditions} is absolutely unquestionable. 

The formulas used in the previous paragraph ( $ \epsilon_{r }  \simeq T_{r}^4 \simeq 1/R^4 ) $ are the standard ones for an adiabatic black body expansion.  However , as it is customary, {\it they do not take into account  the gravitational self energy of radiation. }

Indeed if we consider e.g. a black body of $1/m^3$ (one cubic meter) at room temperature $ kT \simeq 3  \times 10^{-2} eV$ , its total energy is $(kT)^4 m^3 \simeq 10^{8}  MeV$. Its mass  M is hence $10^{-22}   kg $ and its gravitational self energy, assuming a  uniform distribution , $ G M^2 / R \simeq 10^{-10}  \times 10^{-44} /1 \simeq 10^{-54 } J $, completely negligible with respect to its "bare" energy $M c^2 \simeq 10^{-5}  J $ !

However, by  assuming constancy of the photons number, it is immediate to get the temperature T' (see Eq.( \ref {e'ru}) )
at which, radiation gravitational self energy completely swallows the radiation energy

\begin{equation}\label{tP}
R'= G E'_{r,U}/c^4
\end{equation}

This happens for  $ R' \simeq 10^{23}  m$  at $ 3000 K$ .

{\it Indeed  photons, because of the energy mass equivalence, are not only subject to a gravitational field, but are also obviously,  when sufficiently energetic and/or numerous, the source of a sizable gravitational field.}

In other words in our necessarily daring extrapolation of the theoretical tools from the anthropomorphic scale to these extreme conditions (large distances and microscopic scales) we must necessarily consider {\it a gravitational modification of the black body spectrum which entails, as it will be further commented upon shortly,  that the assumption of the constancy of the photon number has to be abandoned} .  Notice that this does not have to apply to primordial conditions but  already to the  radiation dominated era.

As regards  the dominant matter density at present ($E_{m,U} \simeq 10^{80} m_{p} c^2 \simeq 10^{70} J $ , whence $M_{m,U} \simeq 10^{53} kg$ ))

\begin{equation}\label{p}
E_{m,U}^{G}  \simeq  G M_{m,U} ^2/R_{U} \simeq 10 ^{-10}  \times 10 ^{106} / 10 ^{26} \simeq  10 ^{70} J = O(E_{m,U})
\end{equation}

i.e. a remarkable cancellation.

At the Planck scale the result obtained for "ordinary" black holes (\cite{Christ}) applies for the ultimate one i.e. the Planck quantum black hole

\begin{equation}\label{R_{S}}
E _ {P}=M _{P} c^2 -  G  M _{P}^2 /R_{P} = 0
\end{equation}

{\it Thus the initial  fluctuation that originated our Universe had a total zero energy. } 
 
We are therefore in the presence of a situation where the bulk of the present  (i.e. $\simeq $ matter) energy and the initial Planck energy are zero.

 Whether cancellation  happens for the sum of matter and radiation i.e.  , {\it to all significant digits },  remains open. What one can say is that {\it cancellation,  if complete,  would dispose of   the necessity of invoking dark matter} ( as commented upon also in par. 5).)

But more than that,  even if as regards radiation we can go back in temperature to the point where self energy balances the ordinary black body energy, we have (within the present estimates) already reached that point for matter and {\it  by going backwards in time we would have a  total negative energy ! }.  Therefore at earlier times less baryons should have been present. 

In other words {\it  self-energy can act as a gauge in the reconstruction of the history of the Universe, undisputedly proving that a creation mechanism must have been constantly at work}  ( \cite{feynote}, \cite{hoy}).
 
This together with the observation about black holes provides indeed an intriguing setting, although it does not of course contradict our  belief about baryon number conservation at {\it our}  time-space scales.

This is, in a sense, no wonder if we think e.g. at the "eternal" orbits of gravitational systems : by radiating gravitationally they will be less eternal than they seem to us at our anthropomorphic scale.

In conclusion we have seen that the consideration of self gravitational effects (whose theoretical inclusion is beyond any doubt whereas their  actual evaluation is surely questionable)  strongly hints at a plausible  picture of the evolution of our Universe.

Of course local density fluctuations, due to the negative heat capacity of a gravitational system, would independently allow the formation of structures like stars, galaxies,  etc. . 

 Let also add that the present speculations seem to  render a bit more quantitative Feynman's remark (\cite {Feynman}) about the truly amazing coincidence between the  negative gravitational energy and the total rest mass of the Universe   $M_{U} $

\begin{equation}\label{psi}
 M_{U}   ( 1- GM_{U} / c^2 R_{U}  )   =   0
\end{equation}

making the total energy of the universe  zero. 

Thus the Thirring effect \cite {Thir} as quoted in \cite {Feynman} 
which gives the angular frequency $\omega$ caused by a rotating shell of matter of angular frequency $\omega'$ at large distances in {\it flat space}  at the center (in our case the rotating shell of matter would be the whole Universe causing the swinging pendulum follow the shell) 

\begin{equation}\label{psi}
 \omega = \omega' (GM/R c^2)   
 \end{equation}

seems to find (at this semiquantitative level)  an astonishing confirmation. The rotation of the Foucault pendulum   is indeed relative to (and  hence caused by the matter of) the Universe. 

\

We can hence summarize the results of the present approach as follows :

 i) the consideration of neglected self gravitational effects is essential
 
 ii) they play a crucial role both at a classical (Schwarzschild ) and at a quantum (Planck) level and both for matter and for radiation
 
 iii) the evolution of our Universe from the Planck to the present scale reasonably follows, with energy conservation, probably with something exotic  at the very beginning, followed by a radiation era  and matter dominance nowadays and disposing of the apparent problem of quantum fluctuations, with no need for an unbelievably fine tuning.  In addition the fact that space (whose curvature is determined by the energy density) at extreme scales is essentially flat represents another circumstantial evidence (although not compelling) in favour of the present proposal, as it will be also illustrated in par. 5). 
  
 In this connection one further comment  :   {\it space-time  of our Universe, as originated at the Planck scale, is the most rigid   structure or, in other words, the constant k in Einstein's equations must have remained unaltered ever since \cite{Fang}. Hence the experimental evidence for flatness now, is consistent with  flatness even at the primordial quantum fluctuation}, which is commonly thought to be on the contrary dramatically rugged,  and this seems possible only if its total energy is zero. 
 It is also worth mentioning that the large scale flatness is more satisfactorily brought to terms with the presented picture \cite{Christ} of no curvature for the individual bodies which make up the Universe.

 The obvious objection of why black holes would not continuously pop up in our Universe cannot be satisfactorily answered, although we cannot be sure it does not happen and, in case, at what rate.  One could speculate that existing matter  pressure might   act against them .

\

\section{How black is the night ? }\label{night}

 Let us come to some possible additional constraints on the proposed form of matter density  which might help in enlightening the picture of the cavern we live in \cite{vera}.
 
 A first obvious remark is that the self energy of a gravitational body of mass M with a black hole -like matter density 
  $\rho_{sing} = c^2/G (4 \pi r^2)$

\begin{equation}\label{tP}
E_{G} = - G M^2 /R
\end{equation}

is not very different with respect to the  constant density $\rho = M/(4/3 \pi R^3)$ case,  where

 \begin{equation}\label{tP}
E_{G} = -3/5 G M^2 /R
\end{equation}

Especially at a global cosmological level it would not be the case to make much  fuss about two essentially equivalent numbers (3/5 and 1) .

The point is however that whereas in the proposed scenario  {\it cancellation of the total mass by self gravitational effects holds true for any matter subshell }, in the standard constant density case this does not take place. 

In addition these alternative densities weigh differently luminosities : whereas with constant $\rho$ distant regions contribute as much as near ones, leading to the Olbers' paradox (number of stars in every shell $\simeq r^2 dr $ , times  luminosity  $1/r^2$,   =  constant contribution), $\rho_{sing} $ naturally cuts distant contributions adding to Hubble's effect. 

Whereas up to $\simeq 100 \;  Mpc$ the spongy nature of the Universe seems to be out of question, our position would imply a different  non scaling feature  even in the outer regions, a sort of fractal {\cite{cha}, \cite{man}) behaviour , although of a totally different origin, which contradicts the usual homogeneity assumption (this model does not rule out the Big Bang but would account for a dark night even without it) . It is probably worth recalling that, even without unearthing Ptolemy, we already have a privileged reference system : that of the CMR, so that also homogeneity is not necessarily God-given.

Whether this is the case or our position is just an alternative (simplistic) study scheme (although an interesting piece of information along this line seems to come from the 2dF Galaxy Redshift Survey {\cite{coll}),  we cannot convincingly support. Let us however come to another interesting and immediate consequence of it .

\

\section{Hubble's law and the (accelerated?) expansion of the Universe. }\label{}

In connection with the previous point let us concentrate on the first of the two Friedman's equations, solutions to Einstein's field equation $G_{\mu \nu} = 8 \pi G/c^4 T _{\mu\nu} - \Lambda  g_{\mu \nu} $, the one for the velocity , which reads  for zero curvature (\cite{BOOM} , \cite{WMAP}) in the FLRW metric , assuming  isotropy ,

\begin{equation}\label{vel}
(dR/dt  )^2 = 8 \pi G \rho R^2 +( \Lambda/3) \: c^2 R^2
\end{equation}

\

With the proposed  cancellation of gravitational effects in the energy density  it would simply read

\begin{equation}\label{velquad}
 v^2 =   (\Lambda/3) \:   c^2 R^2
\end{equation}

which  immediately leads to the identification

\begin{equation}\label{tP}
\Lambda_{U} /3   = 1/ R_{U}^2 = H_{0} ^2  /  c^2
\end{equation}

and to the approximate Hubble's law 

\begin{equation}\label{velH}
 v =     c \; R/ R_{U}
\end{equation}

where $H_{0} $ is Hubble's constant as measured today as well as $\Lambda_{U}$.

{\it Thus the critical gravitational  density  $\rho_{G}  $ which is thought to govern  our fate would play no role and would not be  determined by Hubble's constant either.}   

 Clearly the necessity of a cosmological constant with our position about gravitational cancellation is manifest : otherwise we would have a zero 
 expansion velocity (notice also that unlike frequently done, the cosmological constant, which might be interpreted as a vacuum energy contribution, has not been reabsorbed into the energy momentum tensor).

 At the Planck scale one would have

\begin{equation}\label{tP}
v^2_{P} =  ( \Lambda_{P}/3) c^2 R_{P}^2  =  H_{P}^2 R_{P}^2
\end{equation}

\

both results, expressing energy conservation,  being summarized as 

\begin{equation}\label{tP}
v^2-   c^2 R^2 / R_{U}^2 = 0 = v^2_{P} -  ( \Lambda_{P}/3) c^2 R_{P}^2  
\end{equation}

One would thus have

\begin{equation}\label{HL}
 H_{0}/H_{P} = \sqrt { \Lambda_{U}/ \Lambda _{P} }=   R_{P}/R_{U} \simeq 10 ^ {-61} = t_{P}/t_{U}
\end{equation}

which can also be cast in the equivalent form

\begin{equation}\label{HL}
R_{U}/t_{U} =  R_{P} / t_{P}= c
\end{equation}

One can  manipulate of Eq. (\ref{RP}) by identifying $\hbar / R_{P} $ with the  momentum;  an  hand-waving prediction of an expansion of the Planck bubble at velocity $c= E_{P}/ p_{P}$ would follow.

The expansion of (the edge of) the causally connected Universe would thus proceed with constant velocity c.  

Thus the proposed scheme, based on physical grounds, is the most economical one which does not seem to contradict present experimental data (as well as the age of the Universe) and with the Planck scenario.

 Does this contradict the present belief of an accelerating Universe \cite{Perl} ?  

If one looks at Fig. 3 of the quoted reference, one clearly sees that the "absurd" (in traditional terms) empty cosmos case is the culprit. Data are accounted for with zero "total" matter density , which has a very clear physical origin in the present approach, and even better with a non zero $\Lambda$ .

Therefore 
 Perlmutter's (\cite{Perl}) valid objection to the standard
 picture that it appears  " a remarkable and implausible coincidence that the mass density, just in the present epoch, is within a factor of 2 of the vacuum energy density  " 
 is overcome by the present consideration of self gravitational effects and
  by assigning a different role to $ \Lambda$ , constant in R but not in time.

Thus the "cosmological constant problem" i.e. the question of why $\Lambda$ is so smaller ($\simeq 10^{-120)}$ than the expected dimensional estimate $M_{P}^4$ appears to be ill posed. Taking into account self energy, total energy disappears and $\Lambda$ enters as an additional parameter whose  role is tied  by QM to the expansion of  the Universe .

Theoretically the prediction of a possible acceleration is governed by the second Friedman's equation which depends crucially on the equation of state parameters (for matter, radiation and so on)
$
w = p/\rho $ , 
therefore heavily model dependent and by no means conclusive.

Even if Eq. (\ref{vel}) might  resemble the corresponding one in   the de Sitter universe   {\cite{de}}, \cite{refe} , the bases   and the results of the present approach are completely different.

First  the reason for the neglect of ordinary matter is argued  , for all times, and not just assumed ;  moreover it is not due to a cancellation between non zero matter density and positive space curvature.  In that sense this might be interpreted as isotropy and homogeneity throughout space (contrary to standard position according to which homogeneity  implies  constant $\rho$ ) but not time.

Second the cosmological constant is given a  time dependence {\cite {wood}}. Indeed a constant ${\Lambda}$ is an essential ingredient in de Sitter's approach and this results in an exponential expansion, which accounts for (the supposed ) early inflation but which surely runs into problems for later times.

Some additional  questions generally arise at this point  about what  has happened before the big bang, what exists  outside the horizon,  etc. etc. . It only seems to me that our Universe is and would keep expanding  , irrespective of whether "before" it had suffered a crunch (if that question makes sense at all, nothing about it being {\it measurable}) . In that case there would have been no singularity 
  due to  the uncertainty principle which intervenes for something  "confined" at the Planck level  acting against its collapse.    

(This is the obvious analogue of e.g. the hydrogen atom. Without the uncertainty principle which furnishes a repulsive term which goes as $1/r^2$  against the attractive Coulomb term which goes as 1/r, the stability would remain unexplained, although no repulsive dark energy has ever been invoked for that).  

So in retrospect, even if introduced for the opposite purpose,  the cosmological constant is everything but  a blunder.
Thus Einstein's question  of why the Universe does not collapse under its own gravitation can be, in a sense, reversed : the Universe, originally constrained at the Planck scale, keeps on expanding on the whole because of the repulsive effect of the uncertainty principle.

As well known, for
the subsystems of the  Universe, i.e clusters,  a separate treatment is needed.

The previous evolution equations apply only on a large scale. Local concentrations do not experience the expansion of the Universe : in addition, as underlined before, the time reconstruction is questionable. Hence whether and how much 
the other additional data from baryonic acoustic oscillations (BAO) might help in constraining {\cite{matt}  $\Omega_{m}$ and $\Omega_{\Lambda}$ appears questionable.

In addition the fact  that one might need unobserved (dark) superstars to explain what is observed  (dark matter with the same $\rho \simeq 1/ r^2$ dependence  but with a markedly different coefficient, to reproduce  $v \simeq constant$ for orbiting stars in the galaxies outskirts) is decoupled from what happens (or not) at a cosmological level. 

It goes anyway without saying that, because of the previous considerations, the virial theorem is more or less violated.
This should hold true even more in the case of supergalaxies which  might not be necessarily  bound  (\cite{van}).

The necessary more detailed analysis is outside the scope of the present work.

\

\section{Conclusions}\label{con}

We have proposed an  alternative cosmological scheme. The central issue of  energy and particles-photons number conservation has been addressed.  It has been shown that the role of self energy is paramount in providing a scenario, that in spite of some daring speculations, seems to be  consistent. Particle creation seems to be an essential ingredient in the evolution of the Universe.

We have shown that the picture that gravity curves space may be abandoned in favour of flatness both locally and  at the cosmological level and substituted by the hypothesis
we might be living   inside a causal connected  expanding black hole.
Future experimental data of this only accessible and hence existing Universe might provide a necessary scrutiny of the proposed theoretical framework.

\

ACKNOWLEDGMENTS

\

I wish to thank  M. Lucchesi for the  "naif " remark which prompted this investigation and P.G. Prada Moroni, G. Morchio and M. Velli for useful discussions and for a critical reading of the manuscript  .


\begin{thebibliography}{999}
\bibitem{feynote}
 R.P. Feynman Lectures on gravitation. Penguin books
  " It is exciting to think that it costs nothing to create a new particle, since we can create it at the center of the Universe where it will have a negative gravitational energy equal to $Mc^2$. "
\end{thebibliography}
\end {document}